\documentclass[preprint, amsmath,amssymb, aps]{revtex4-2}
\usepackage{chessfss,graphicx}
\usepackage{dcolumn}
\usepackage{bm}
\usepackage{xcolor}
\usepackage{lineno}
\usepackage{ulem}
\newcommand{\avg}[1]{\left\langle #1  \right\rangle}

\begin{document}

\title{Emergent Complexity in the Decision-Making Process of Chess Players}

\author{Andr\'es Chacoma* (achacoma@df.uba.ar)}
\affiliation{Instituto de F\'isica Interdisciplinaria y Aplicada, CONICET - Universidad de Buenos Aires, Argentina}
\affiliation{Departamento de F\'isica, Facultad de Ciencias Exactas y Naturales, Universidad de Buenos Aires, Argentina.}

\author{Orlando V. Billoni}
\affiliation{Universidad Nacional de Córdoba. Facultad de Matemática, Astronomía, Física y Computación. Grupo de Teoría de la Materia Condensada. Córdoba, Argentina}
\affiliation{Consejo Nacional de Investigaciones Científicas y Técnicas, CONICET, IFEG. Córdoba, Argentina.}

\begin{abstract}

In this article, we study the decision-making process of chess players by using a chess engine to evaluate the moves across different pools of games.
We quantified the decisiveness of each move during the games using a metric derived from the engine's evaluation of the positions. We then performed a comparative analysis across players of varying competitive levels.
Firstly, we observed that players face a wide spectrum of the decisiveness metric, evidencing the complexity of the process. By examining groups of winning and losing players, we found evidence where a decrease in complexity may be associated with a drop in players' performance levels.
Secondly, we observed that players' accuracy increases in positions with high values of the decisiveness metric regardless of competitive level. Complementing this information with a null model where players make completely random legal moves allowed us to characterize the decision-making process under the simple strategy of making moves that minimize the decisiveness metric. 
Finally, based on this idea, we proposed a simple model that approximately replicates the global emergent properties of the system.
\end{abstract}

\maketitle

\section{Introduction}

Within the framework of complexity science, a sports competition can be thought of as a complex system whose dynamics evolve based on interactions of cooperation and competition, modulating players' behavior in response to highly changing environments.
The inherently complex nature of sports competitions, a defining feature that has long fascinated academics, has seen a surge in scholarly interest in recent years
~\cite{ribeiro2012anomalous,clauset2015safe,kiley2016game,ibanez2018relative,ruth2020dodge,martinez2020spatial,yamamoto2021preferential,chacoma2020modeling, chacoma2021stochastic, chacoma2022complexity,chacoma2025identification,chacoma2022simple, chacoma2023probabilistic}.
In this context, chess has historically been one of the most studied competitive domains. It has accompanied human history since ancient times~\cite{prost2012impact}, served as a benchmark for computer evolution in the last century~\cite{shannon1950xxii}, and more recently, driven breakthroughs in artificial intelligence~\cite{mcgrath2022acquisition}.
Chess offers utility beyond mere competition, serving as a valuable tool for assessing diverse human capabilities. For instance, it has been employed to study cognitive performance in both professional and novice players \cite{Duan12PO}, decision-making processes~\cite{Sigman10FN, Leone17C}, the relationship between expertise and knowledge \cite{Chassy11PO, Sala2017}, and even gender disparities in competitive settings \cite{brancaccio2023scientific}.

Chess derives its utility as a research tool from two key pillars: complexity and popularity. In the last years, quantitative studies have characterized chess's complexity by analyzing collective player behavior through large-scale datasets~\cite{Blasius09PRL,Ribeiro13PO, Perotti13EPL,Schaigorodsky14PA, Atashpendar16EPL, Schaigorodsky16PO, 
almeira2017structure, demarzo2023quantifying, barthelemy2023statistical}.
Blasius and T\"{o}njes \cite{Blasius09PRL} reported a Zipf's law behavior for the pooled distribution of chess
opening weights and these findings were explained analytically as a multiplicative process.
Studying the growing dynamics of the game tree, Perotti et al. have found that the emerging Zipf and Heaps laws can be explained
in terms of nested Yule-Simon preferential growth processes~\cite{Perotti13EPL}. 
Ribeiro et al.~\cite{Ribeiro13PO} investigated the move-by-move dynamics of the white player's advantage finding
that the diffusion of the move dependence of the White player’s advantage is non-Gaussian, has long-ranged
anti-correlations and that it becomes super-diffusive after an initial period with no diffusion.
Recently, Barthelemy~\cite{barthelemy2023statistical} have used Stockfish to search critical points~\cite{dorfman2001method},
performing a statistical analysis of the difference between the evaluations of the best move and the second best move introducing the parameter $\Delta$.

In this paper, we first review and validate statistical results on chess gameplay from the literature~\cite{Blasius09PRL,barthelemy2023statistical}, while introducing new observables reported here for the first time. Specifically, we propose a null reference model to identify key game-level characteristics.
These results enable us to develop a statistical framework for chess games. The model incorporates microscopic mechanisms of player decision-making during gameplay, explaining the emergence of macroscopic statistical patterns.
To implement this approach, we performed a data-driven analysis of chess games using the Stockfish engine. This tool enables us to quantify how player decisions influence game dynamics.

The manuscript is organized as follows: Section II contextualizes our findings through a review of recent literature. Section III details the datasets and acquisition methodology. Section IV presents our analytical results, and Section V concludes with a summary of key contributions.

\section{Related literature }

In recent years, breakthroughs in data acquisition technologies—such as high-frequency tracking systems, wearable sensors, and advanced computer vision—coupled with unprecedented computational power, have revolutionized the quantitative study of sports. These innovations have not only intensified academic interest but also enabled rigorous multidisciplinary research into the emergent properties of competitive sports~\cite{ribeiro2012anomalous,clauset2015safe,kiley2016game,ibanez2018relative,ruth2020dodge,martinez2020spatial,yamamoto2021preferential}.
In this context, we have previously analyzed sports
such as football  \cite{chacoma2020modeling, chacoma2021stochastic, chacoma2022complexity},
volleyball \cite{chacoma2022simple}, and paddle tennis \cite{chacoma2023probabilistic}.
With the same spirit, this study focuses on the game of chess.

Today, chess enjoys global popularity, supported by large and active communities of players across all skill levels. Competitive matches take place both traditionally over-the-board and through online platforms, generating vast quantities of game records, which offers unparalleled opportunities for large-scale quantitative analyses.
In this context, the expertise of each player, is statistically well characterized through the Elo system, introduced by the physicist Arpad Elo~\cite{elo2008rating,Glickman95ACJ}. 
This rating system also exhibits predictive power, as it enables the estimation of match outcomes between two ranked players.

In what follows, we provide a concise review of key chess-related studies to contextualize the extensive research landscape surrounding this game.
Through systematic analysis of timestamped chess games, Schaigorodsky et al. identified persistent long-range memory effects in game dynamics~\cite{Schaigorodsky14PA, Schaigorodsky16PO} and correlations between communities of players and their level of expertise~\cite{almeira2017structure}.
Atashpendar et al.~\cite{Atashpendar16EPL} have examined chess as a complex system, specifically analyzing the structure of its state space. Their analysis revealed that the state space comprises multiple distinct clusters with infrequent inter-cluster transitions, and that skilled players only partially explore certain regions of this configuration space.
In a multidisciplinary approach, De Marzo and Servedio~\cite{demarzo2023quantifying} adapted the Economic Fitness and Complexity algorithm to quantify both opening difficulty and player skill levels in chess.
Leveraging AI-powered analysis, Adhikari et al.~\cite{Adhikari2024mistakes} demonstrated that most in-game errors induce tilt, subsequently degrading player accuracy. However, their findings reveal an intriguing exception: following serious blunders, players exhibit increased discipline in subsequent moves.
In a related approach, Anbarci et al.~\cite{anbarci2024ai} developed an AI-based framework for objective tiebreaking in chess. Their methodology assesses move quality by quantifying the deviation between players' actual moves and engine-derived optimal moves.
Additional studies have examined structural aspects of chess tournaments, including pairing algorithms and participant self-selection. Csató~\cite{csato2024most} demonstrated that most Swiss-system tournaments exhibit significant fairness issues, while Linnemer~\cite{linnemer2016self} analyzed how player self-selection biases tournament outcomes.
Within cognitive science research, Gonzalez et al.~\cite{gonzalez2016cognitive} employed chess as an experimental paradigm to investigate cognitive performance under competitive conditions, demonstrating significant influences of psychological factors on decision-making processes. Complementing this perspective, Aldous~\cite{Aldous2017Elo} conducted a mathematical analysis of the Elo rating systems, developing improved algorithms for tracking players' evolving skill levels.

\section{Data}

In this paper, we analyze three public datasets, with the aim of conducting an analysis based on the competitive level of the players. 
The first dataset, referred to as FIDE, contains $3000$ games from the World Championships and European Championships from the years 2019, 2020, 2021, and 2022. These games are of a high competitive level.
The second dataset, referred to as LC high Elo, contains $1000$ games played by players on the {\it Lichess} platform \cite{weblichess} where the average rating between the two competitors is above 2500. The rating used here is the one assigned by the Lichess portal, calculated based on the player's game history on that platform. The games collected in this dataset correspond to an intermediate competitive level.
Finally, the dataset referred to as LC low Elo contains $3000$ games played by players on the Lichess platform where the average rating between competitors is below 1500. These games are from beginner players, with a low competitive level.
All datasets contain classical chess games in portable game notation (pgn) format.
The games associated with the FIDE dataset were downloaded from the website {\it The Week in Chess} \cite{webtwic}, and the others from the {\it Lichess} database \cite{lichessdatabase}.

\section{Results and discussion}

\subsection{Global emergents in chess}

When a chess player faces a position on the board, they must evaluate it, decide which move is the most advantageous, and execute it. This decision-making process is based on the game's tactical and strategic principles, which, like in any sport, the player acquires and refines during their training.
A chess game, therefore, can be thought of as a succession of positions that both players must evaluate accurately until the end of the game.
In each of these positions,  a finite number of legal moves, $\mu$, define the range of possible plays for the competitor.
Since $\mu$ is a finite random variable, the evolution of the game can be represented as a walk in a vast and complex decision tree.

\begin{figure}[t!]
\centering
\includegraphics[width=1.\textwidth]{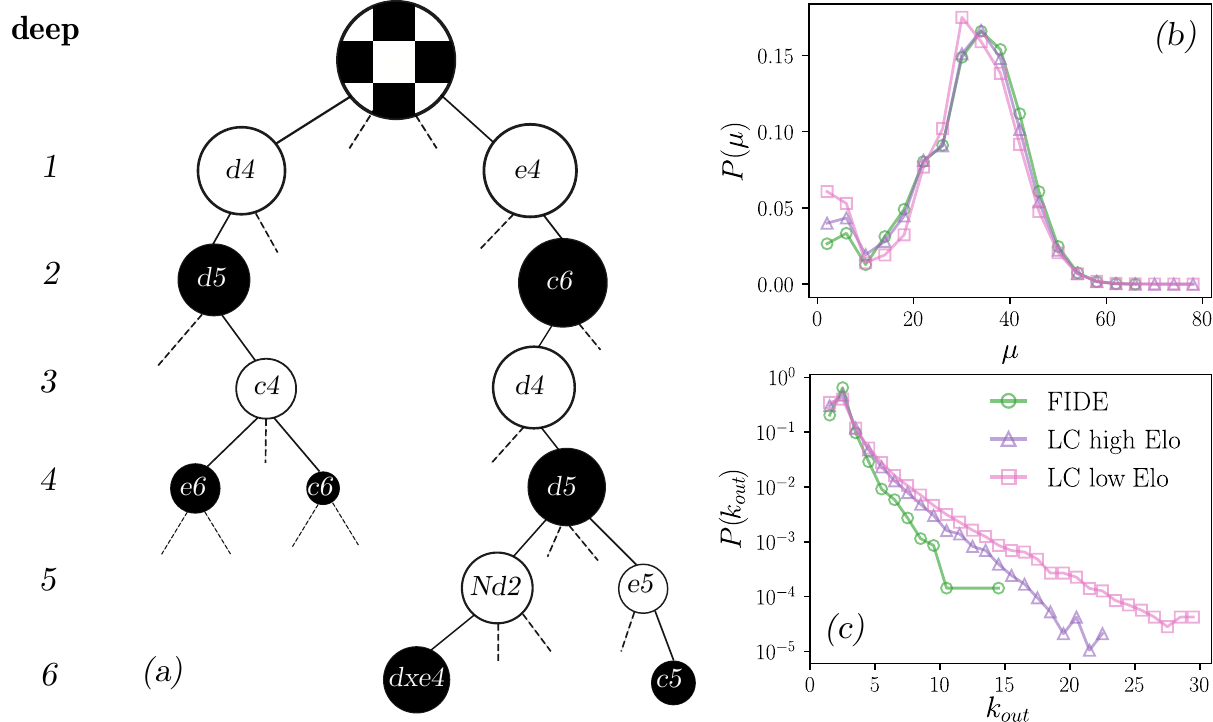}
\caption{
The Chess Tree.
(a) Schematic of a small portion of the chess tree. Each node represents a board position, with the label indicating the last move made. The size of the nodes (not to scale) indicates that some nodes are more popular than others.
(b) At each turn, depending on the stage of the game, players have a certain number of legal moves, \( \mu \). The plot shows the distribution \( P(\mu) \).
(c) The empirical tree, constructed from the data, can be thought of as a directed complex network, where the nodes have \( k_{out} \) outgoing links. The plot shows the distribution \( P(k_{out}) \).
}
\label{fi:arbol}
\end{figure}

In Fig.~\ref{fi:arbol}~(a), we show a non-scaled visualization of the decision tree structure, highlighting the initial moves of the Queen’s Gambit opening (left) and the Caro-Kann Defense (right).  
This figure is a schematic representation intended to illustrate the existence of different paths, the presence of more frequently visited nodes at various levels, and the overall complex structure of the tree.  
The size of the circles in the graph represents the popularity of each node (not to scale), that is, the number of times, $n$, that position was observed in the database.  
From now on, we will refer to the number of moves as the depth, $d$, corresponding to the depth of the tree.
In Fig.~\ref{fi:arbol}~(b), we show the distribution $P(\mu)$. We calculated the mean value and the standard deviation in the three cases and obtained similar results, $\avg{\mu}\approx30$ and $\sigma_{\mu}\approx12$.
We can see that the distribution shows an approximately normal behavior with a large peak around $\mu\approx35$ and a smaller peak around $\mu\approx5$. The latter is associated with check situations, where legal moves are reduced. Note that lower-level players seem to be more exposed to these situations. 
This may be due to the fact that high-level players tend to abandon the game before reaching those stages.
Notably, these results are similar to those reported in \cite{barthelemy2023statistical} for the case of engine chess games, indicating that the approximately normal shape of $P(\mu)$ is something intrinsic to the game.
Lastly, we constructed a directed network with nodes that meet $n>1$ and studied the degree distribution, $P(k)$, which we show in Fig.~\ref{fi:arbol}~(c). We call this network the empirical tree because it is formed only by the nodes that appear in the database. In all three cases, we observed curves with a sample mean of $\avg{k}\approx1$. This indicates that the tree is mainly formed by chains. Interestingly, the variance decreases as the level of the players increases. Higher-level players tend to explore fewer variants compared to lower-level players, probably because they have a greater ability to decide primarily on good lines.

\begin{figure}[t!]
\centering
\includegraphics[width=1.\textwidth]{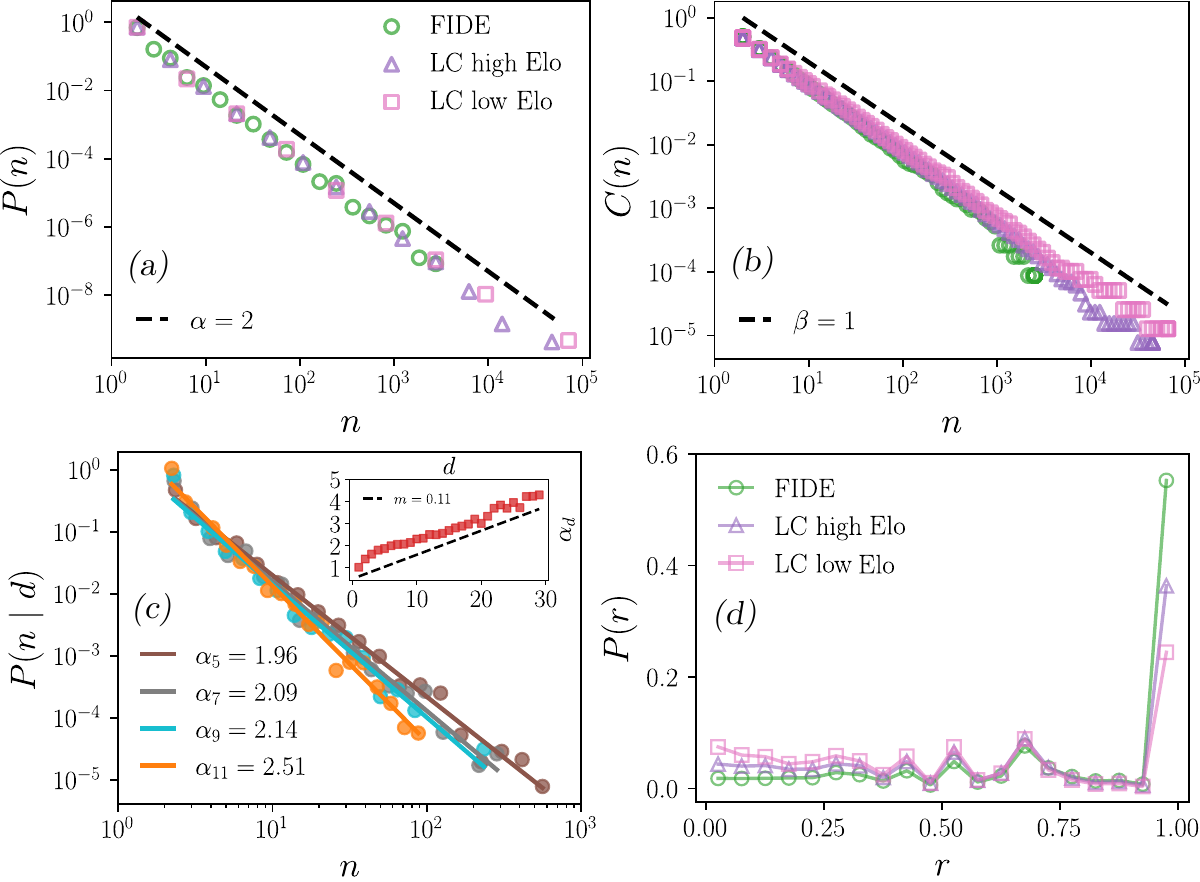}
\caption{
Global Emergent Properties of the System.
(a) Distribution of node popularity, \( P(n) \), in the three datasets analyzed: FIDE, LC high ELO, and LC low ELO. Dashed lines show the theoretical distribution \( P(n) = n^{-2} \).
(b) Complementary cumulative distribution of node popularity, \( C(n) \), in the three datasets analyzed. Dashed lines show the theoretical distribution \( C(n) = n^{-1} \) (Zipf's law).
(c) Conditional popularity distribution at a specific tree depth, \( P(n|d) \). Solid lines show a nonlinear fit to the curves used to calculate the exponents \( \alpha_d \). The inset shows the relationship between \( \alpha_d \) and depth \( d \).
(d) Distribution of ratios, \( P(r) \), for the three datasets analyzed.
}
\label{fi:emergentes}
\end{figure}

In Fig.~\ref{fi:emergentes}~(a) and (b), we show the distribution of node popularity, $P(n)$, and the complementary cumulative distribution, $C(n)=1-P(N \leq n)$. It is well known that the game tree is self-similar and $C(n)$ follows Zipf's law \cite{Blasius09PRL}. In dashed lines, we show $P(n)\propto n^{-\alpha}$, and $C(n)\propto n^{-\beta}$, with $\alpha=2$ and $\beta=1$. 
Notably, the distributions do not appear to depend on the competitive level of the players.
By collecting $N=10^4$ random games from the three datasets, we measured the node popularity at different depth levels to obtain $P(n|d)$. 
In Fig.~\ref{fi:emergentes}~(c), we can see that these distributions also follow a power law, $P(n|d)\propto n^{-\alpha_d}$. Likewise, the exponents $\alpha_d$ follow a linear trend as a function of depth, as shown in the inset of the figure. This trend can be estimated as $m=1/\log(N)\approx 0.11$ \cite{Blasius09PRL}. The dashed lines show how the theoretical line, $\alpha_d \propto m \times d$, closely approximates the empirical trend.

Finally, we define the branching ratio, $r$, as follows: Let $i$ and $j$ be two connected nodes in the network located at levels $d$ and $d+1$. Let $n^i_{d}$ and $n^j_{d+1}$ be the popularities of these nodes, then $r^{ij}=n^j_{d+1}/n^i_d$. Note that if $r^{ij}=1$, all players who passed through node $i$ chose the same move that leads to node $j$. Conversely, if $r^{ij}\approx0$, few players who passed through node $i$ chose the move leading to node $j$. In Fig.~\ref{fi:emergentes}~(d), we show the distribution $P(r)$ for the three analyzed datasets. 
To calculate $r^{ij}$ for each pair of nodes, we require that $n^j_{d}>1$ and $n^j_{d+1}>1$. 
We can see the presence of a peak around $r\approx1$. For the rest of the range, the distribution approximates a uniform distribution. These observations coincide with those reported in \cite{Blasius09PRL}. 
It is also observed that an increase in the competitive level of the players produces a larger peak around $r\approx1$ and a lower values in the distribution for $r<0.5$. 
This is consistent with what is observed in Fig.~\ref{fi:arbol}~(c), where lower-level players tend to explore more paths than higher-level players.

\subsection{Quantifying the Spectrum of Decisions}

During the game, players must evaluate different types of positions to decide which is the best move to make. A natural question is whether all decisions have the same impact on the game. In this section, we discuss this issue.

To quantify the players' decisions, we used one of the best chess engines available today: StockFish 16 (SF) \cite{SF}. 
For our study, Stockfish (SF) was configured to operate at its maximum skill level, with an approximate Elo rating of $3500$. Note, that the Elo difference between super-grandmasters (approximately 2700–2800 Elo) and SF in this setting is estimated to be between 600 and 700 Elo points. According to the logistic expression for a player's expected score~\cite{elo2008rating}—which combines the probability of winning with half the probability of drawing—the probability of a human player scoring against SF lies between 0.0174 and 0.03. This implies that, over a set of 100 games, a human player would obtain approximately 2 to 3 points. In this context, it is clear that SF significantly outperforms human capabilities. 
Additionally, given that computation times can be very long for certain board positions, we limited the engine's search depth to 20 moves. This configuration allows SF to maintain superhuman performance~\cite{ferreira2013impact}, while also optimizing the computation time required to analyze our extensive dataset.
It is important to note that this limitation may reduce the engine’s performance in specific board positions—for example, in complex endgames such as the case king, bishop, and knight versus king. However, such cases are rare and can be considered negligible within the broader context of the analyzed games.

\begin{figure}[t!]
\centering
\includegraphics[width=1.\textwidth]{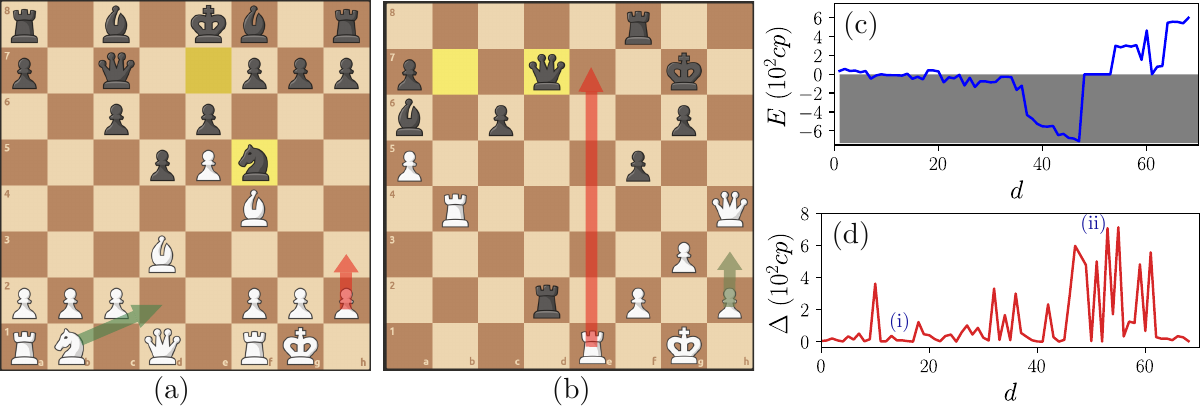}
\caption{
Evaluation and Quantification of Decisions.
(a) Example of a position where the decision of the moving player (White) is not very decisive for the game.
(b) Example of a position where the decision is decisive for the game.
(c) StockFish evaluation of an anonymous game from the LC low Elo dataset.
(d) Value of \( \Delta \) at each position in the game mentioned in the previous panel. References (i) and (ii) indicate the moments in the game where the positions shown in panels (a) and (b) appear.
}
\label{fi:tableros}
\end{figure}

Given a position, SF evaluates it and quantifies its evaluation with a number we call $E$. In Fig.~\ref{fi:tableros}~(c), we show the evaluation curve of all positions in a game, $E$ vs. $d$. In this scheme, if $E\approx0$, SF indicates that the game is even. If $E>0$ ($E<0$) the advantage is for the player with the White (Black) pieces. 
The values of $E$ are given in units of centipawns, $[E]=cp$, where each unit represents 1/100 of a pawn advantage.
Given a position, SF also allows us to rank all possible moves, from best to worst, based on the evaluation it would give to the resulting position if we made that move.
In Fig.~\ref{fi:tableros}~(a) and (b), we show two positions from a game, both with White to move. In each case, we used SF to obtain the top three moves (M). 
For the position in panel (a), we have: 
(1) $M_1=h3$ with $E_1=+0.37$, 
(2) $M_2=\symknight d2$ with $E_2=+0.37$, and 
(3) $M_3=\symrook e1$ with $E_3=+0.35$. 
For the position in panel (b), we obtain:
(1) $M_1=\symrook e7$ with $E_1=+2.96$, 
(2) $M_2=h3$ with $E_2=-4.11$, and 
(3) $M_3=\symrook bb1$ with $E_3=-4.46$. 
All $E$ values are expressed in $10^2cp$.

Note that in the case of the position in panel (a), the decision that the player makes does not seem to matter much.
Any of the three moves SF suggested results in a similar evaluation of the game. We say that this type of position is ``not very decisive'' because a bad decision cannot significantly alter the course of the game. Of course, we are talking about rational decisions, excluding irrational moves like $\symbishop h6$, a move that sacrifices a bishop for no reason.
Conversely, in the position in panel (b), it is very important that the player makes the correct move. In this position, if White plays $M_1=\symrook e7$, it executes a fork, attacking both the king and queen with the rook simultaneously. Being in check, the black king must move, and White will capture the black queen, which gives a decisive advantage.
Note that only $M_1$ gives an advantage to the player with White; any other option will put them at a disadvantage. We say that this type of position is ``highly decisive'' because a bad decision can change the course of the game.

To quantify how decisive the position the players are facing is, previous works have proposed measuring the parameter $\Delta= |E_1 - E_2|$ \cite{barthelemy2023statistical}. Note that if $\Delta$ is small, it indicates a less decisive position, and if $\Delta$ is large, it indicates a highly decisive position, such as those shown in Fig.~\ref{fi:tableros}~(a) and (b), respectively. In particular, if $\Delta > \avg{\Delta}$, we will say that the position is a ``tipping point'', because what the player decides in that position can be crucial for the development of the game.
For the game from which we extracted the positions in Fig.~\ref{fi:tableros}~(a) and (b), we calculated the relation $\Delta$ vs. $d$ shown in Fig.~\ref{fi:tableros}~(d). Note how at the beginning of the game, the values of $\Delta$ remain small, but as the players progress in the game, the values of $\Delta$ begin to indicate the presence of more decisive positions.

It is interesting to note that if SF were a perfect evaluating machine, meaning it could compute the chess tree to an infinite depth, the evaluation would only admit three values: $E=-E_{max}$, $0$, $+E_{max}$. These values would indicate whether the player is in a position where Black wins, a draw, or White wins, respectively.
In this case, the gap would also only admit three values: $\Delta=0$, $E_{max}$, $2E_{max}$, indicating that the position is not decisive, somewhat decisive, or very decisive, respectively.
Since SF is not a perfect evaluating machine, the evaluation values $E$ can be thought of as proportional to a probability of winning, drawing, or losing. In this framework, the inherent limitations in SF's calculation connect better with human reality. The information it provides is more representative of the players' analytical power compared to what a perfect evaluating machine would provide.

As we can see, the results suggest that $\Delta$ can be an interesting proxy to quantify the spectrum of decisions to which players are subjected, depending on how decisive these are for the game.
In the next section, we statistically characterize this parameter.

\subsection{Emergence of Complexity in the Decision-Making Process}

\begin{figure}[t!]
\centering
\includegraphics[width=1.\textwidth]{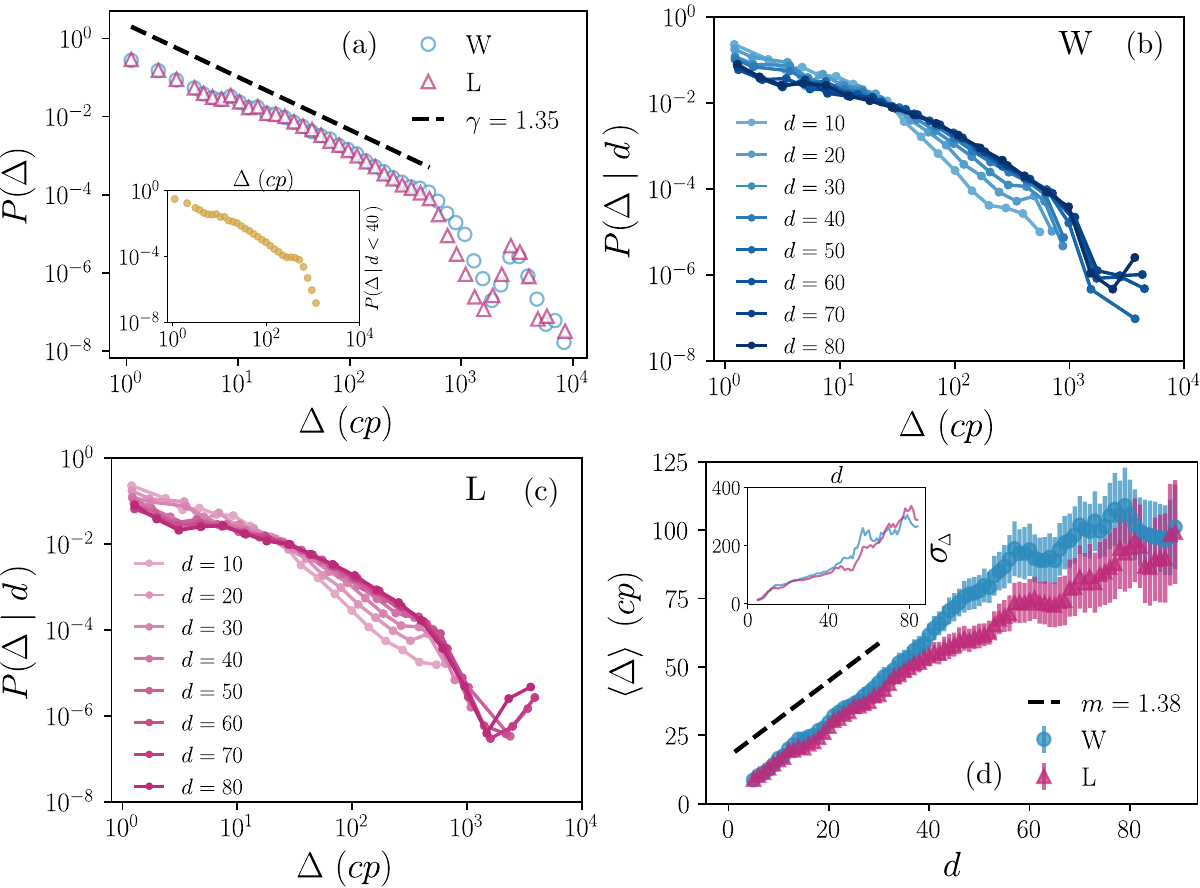}
\caption{
Complexity in the Decision-Making Process.
(a) Distribution \( P(\Delta) \) in groups of winning players (W) and losing players (L). Dashed lines show the theoretical distribution \( P(\Delta) = C\Delta^{-\gamma} \). The inset shows the same probability distribution for \( \Delta \) values observed at depths \( d < 40 \).
(b) and (c) show the distributions \( P(\Delta|d) \) for different values of \( d \) in W and L groups, respectively.
(d) Mean value of \( \Delta \) as a function of depth for W and L groups. The inset shows the standard deviation, \( \sigma_{\Delta} \).
}
\label{fi:comp_1}
\end{figure}

For this particular analysis, we used the FIDE dataset. First, we calculated the $\Delta$ vs. $d$ curves for all games and separated the dataset into two groups: in group W, we have the $\Delta$ values associated with the players who won the game, and in group L, those who lost. In this analysis, we excluded draws; we are interested in studying only decisive games.
In Fig.~\ref{fi:comp_1}~(a), we show the distribution $P(\Delta)$ for groups W and L. We can see a power-law trend over three orders of magnitude with a cutoff at $\Delta\approx1000$. In that range, using the maximum likelihood method \cite{clauset2009power}, we fit the curve associated with group W to the expression $P(\Delta)= C \Delta^{-\gamma}$, obtaining the value $\gamma=1.35$. The fact that the distribution is power-law-like shows that players are subjected to a heterogeneous landscape of decisions during the game. This is a clear indication of the emergence of complexity in the system. In the inset, we show the distribution for all the moves such $d<40$, $P(\Delta|d<40)$. For this plot, we took all $\Delta$ values from both groups. We can see that the small peak at $\Delta\approx5000$ in the main figure disappears, indicating that these extreme values manifest only in the defining stage of the game.

This results show that $\Delta$ can take values that strongly depend on the stage of the game and, consequently, on the depth of the tree. To explore this further, we decided to study the distributions $P(\Delta|d)$ shown in Figs.~\ref{fi:comp_1}~(b) and (c). For both groups, we can see that as $d$ increases the probability of finding high values of $\Delta$ increases, while the probability of finding low values decreases. It is interesting that, regardless of the depth, the distributions appear to have a heavy tail, indicating that the complexity in the decision-making process persists throughout all stages of the game.

To complement this analysis, in Fig.~\ref{fi:comp_1}~(d), we present the evolution of the mean value $\avg{\Delta}$ as a function of depth $d$. The bars indicate the $99\%$ confidence interval around the estimator at a given depth. We observe an approximately linear evolution up to $d=40$. In this range, the mean value of $\Delta$ increases at a rate of $m=1.38$. 
As players delve deeper into the game, the game simplifies, and the analysis becomes more accurate. Consequently, more decisive positions emerge, and the $\Delta$ values increase.
For $d>40$, we observe that the values of $\avg{\Delta}$ associated with group L begin to decrease. At this stage of the game, the matches start to be decided. Players who are losing begin to run out of tactical resources, and the decisions they make become less important, leading to the observed decrease in $\Delta$ values. 
The inset shows the evolution of the standard deviation estimator, $\sigma_{\Delta}$. We observe an increasing trend as players progress in the game. For $d>40$, we see something similar to what happens with the mean value, as the curve associated with losing players decreases slightly, indicating lower dispersion. 
It is interesting to note that this decrease in the standard deviation can be connected with a decrease in complexity levels compared to the winners' group. In this sense, a loss of complexity in group L may be connected with a decrease in players' performance.

\subsection{Players' Accuracy}

\begin{figure}[t!]
\centering
\includegraphics[width=1.\textwidth]{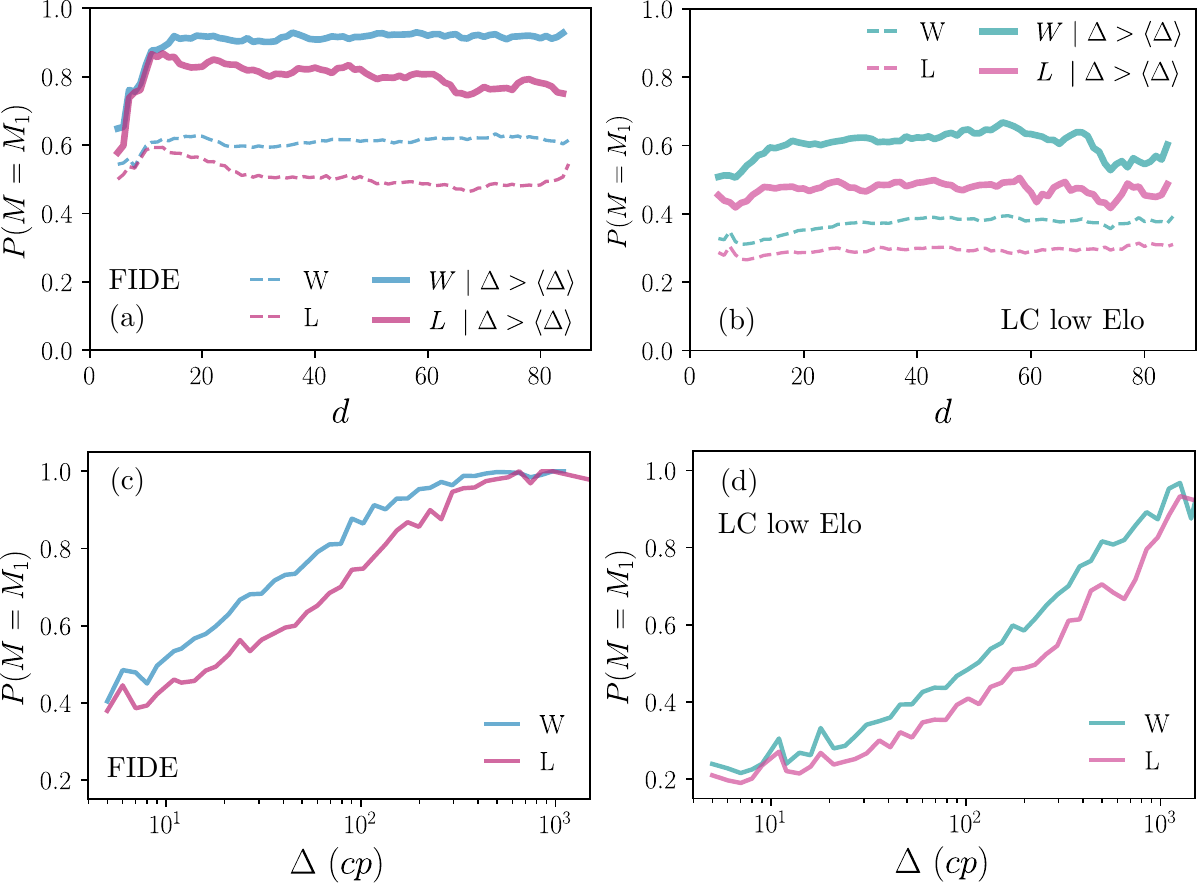}
\caption{
Players' Accuracy.
In (a) and (b), we show the probability that players execute the optimal move as a function of tree depth, $d$, for the FIDE and LC low Elo datasets, respectively. The datasets are separated into the W and L groups.
Dashed lines show the total probability, and solid lines show the probability conditioned on the presence of tipping points.
In (c) and (d), we show the probability that players execute the optimal move as a function of \( \Delta \) value for the FIDE and LC low Elo datasets, respectively. The datasets are separated into the W and L groups.
}
\label{fi:comp_2}
\end{figure}

In this section, we study the accuracy of the players. 
In Fig.~\ref{fi:comp_2}~(a), we show the probability that players in the FIDE database execute the optimal move, $M=M_1$. 
The data is shown as a function of depth $d$ and separated into groups W and L. 
Dashed lines represent the total probability, $P(M=M_1)$,
calculated over all the positions observed at different depth levels.
Solid lines represent the probability conditioned on the presence of tipping points, $P(M=M_1~|~\Delta>\avg{\Delta})$,
calculated over all the positions that exceed the mean value of $\Delta$ of the game to which they belong.
Focusing on the latter, we observe that in both W and L, the probability shows an increasing trend up to $d\approx10$. For $d>10$, in group W, the probability grows and stabilizes at approximately a constant value above $0.9$. On the other hand, in group L, we see that the probability shows a slightly decreasing trend across the entire range. A similar behavior is observed in the total probability, but with significantly lower values. In both cases, the probability associated with W is higher than that of L for all $d$.
In Fig.~\ref{fi:comp_2}~(b), we show the same relationships as in panel (a) but calculated on the LC low Elo dataset. In this case, the competitive level of the players is much lower, so the probability of executing the optimal move decreases significantly compared to elite players (FIDE). Nevertheless, as in the previous case, the probability associated with the winning group W is higher than that of L for all values of $d$.

It is important to note that regardless of the competitive level and the stage of the game, all players improve their performance in the presence of tipping point positions. This highlights a dependence of accuracy on the values of $\Delta$. To further investigate this, we study the probability the players make the best move $P(M=M_1)$ vs. $\Delta$ in the FIDE and LC low Elo datasets. The results are shown in Fig.~\ref{fi:comp_2}~(c) and (d).
In both panels, we can see that the probability increases as $\Delta$ increases. However, the probability for FIDE players seem to increase at a faster rate. As expected, for the same value of $\Delta$, players in group W generally show a higher probability than those in group L in both datasets.

\subsection{Null Model: Random Walk in the Chess Tree}

In this section, we compare some of the results observed in the real system with a null model. The idea is to try to better understand the emergent properties associated with the parameter $\Delta$. Our null model is based on simulations of the chess game where competitors make random legal moves. To perform these simulations, we coded a Python script that simulates random games (RW) using the {\it chess} library as the core. With this script, we simulated 1000 games with a total depth of $D=80$ and then calculated the values of $\Delta$ for each of these games, following the same procedure as with the real games.

\begin{figure}[t!]
\centering
\includegraphics[width=1.\textwidth]{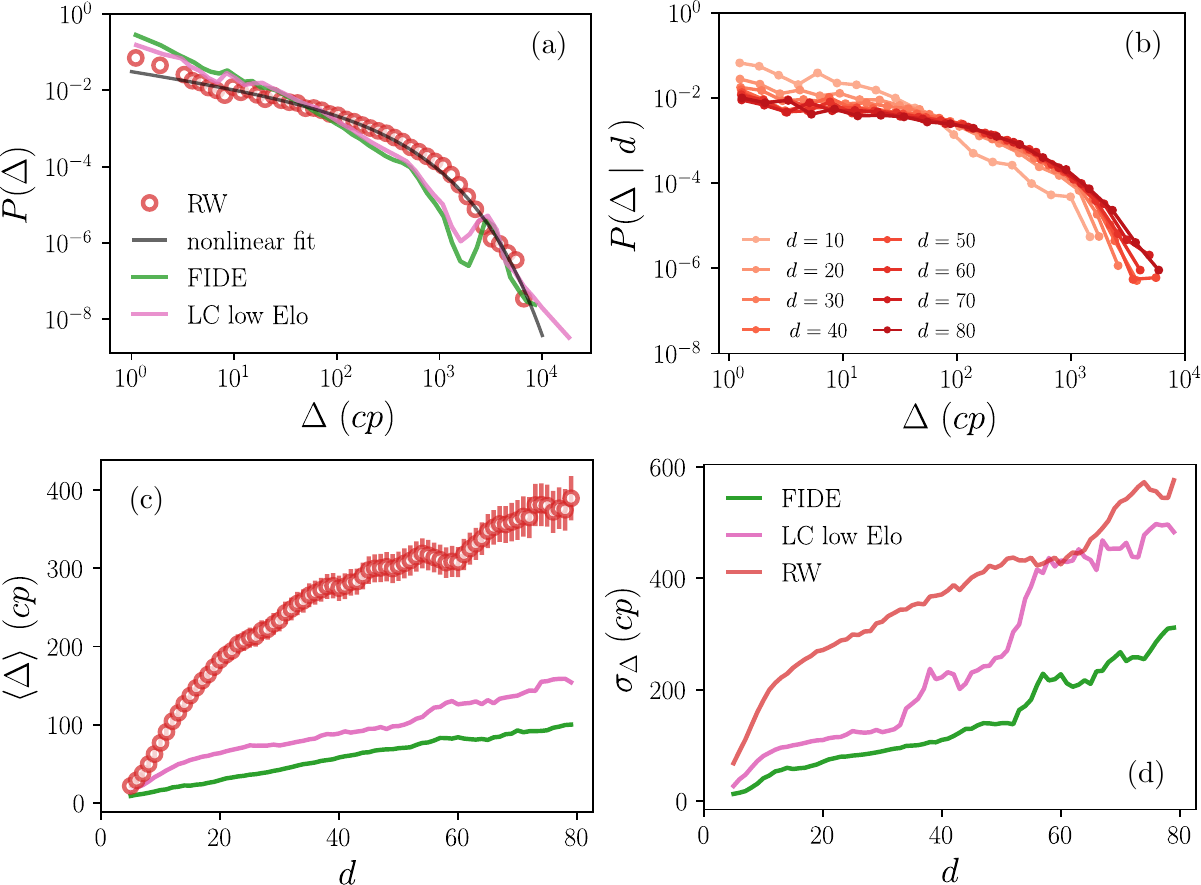}
\caption{
Results of the Null Model (RW).
(a) Probability \( P(\Delta) \) obtained from the null model (RW) compared with the FIDE and LC low Elo datasets.
Solid line indicates the performed nonlinear fit to the distribution
linked to RW data. 
(b) Depth-conditioned distributions. \( P(\Delta|d) \) for the null model case.
(c) Relationship \( \avg{\Delta} \) vs. \( d \) in the null model case, compared with the FIDE and LC low Elo datasets. The colors of the curves are referenced in panel (a).
(d) Relationship \( \sigma_{\Delta} \) vs. \( d \) in the null model case, compared with the FIDE and LC low Elo datasets.
}
\label{fi:rw}
\end{figure}

In Fig.~\ref{fi:rw}(a), we compare the distribution $P(\Delta)$ obtained from the null model with the distributions obtained from the games in the FIDE and LC low Elo datasets. 
The distribution $P(\Delta)$ for the random case, can be fitted with a stretched exponential curve of the form,
\[ 
f(x; \lambda, \beta) = \beta \lambda x^{\beta - 1} e^{-\lambda x^{\beta}}, 
\]
where $x=\Delta$, and $\beta$, $\lambda$ are free parameters.
By performing the nonlinear fit, we obtain $\lambda=0.056$ and $\beta=0.584$, which gives us an excellent agreement, as we can see in the plot.
On the other hand, we can see that in the simulated games, there are significantly higher values of $\Delta$ compared to the real cases. 
Moreover, the curve associated with lower-level players also shows the presence of higher values of $\Delta$ compared to elite players.
These results uncover that the observed values of $\Delta$ may depend on the competitive level of the players.

In Fig.~\ref{fi:rw}~(b), we show the distributions of $P(\Delta|d)$ for the RW case. We can see that the distribution tends to flatten out as $d$ increases. Comparing this with what was obtained for the FIDE dataset in Fig.~\ref{fi:comp_1}~(b) and (c), in RW, the curves are more similar to each other in the interval $100<\Delta<1000$, where for the FIDE case, significant differences are observed.

The values of $\Delta$ in the null model games are much larger at all depths than in the real case. This is clearly evidenced in the plot of $\avg{\Delta}$ vs. $d$ in Fig.~\ref{fi:rw}~(c). In this case, we can see that the curve associated with RW takes values up to four times higher than the curve associated with FIDE. Likewise, the curve associated with lower-level players is significantly higher than that of elite players and much lower than for the null model.
In Fig.~\ref{fi:rw}(d), we show the evolution of the standard deviation estimator, $\sigma_{\Delta}$, as a function of depth. For all values of $d$, we see that the standard deviation increases as the competitive level decreases.

The results indicate that when the level of the players increases, the values of $\avg{\Delta}$, as well as their fluctuations, decrease independently of the game depth. This reduction of $\avg{\Delta}$ has two main effects: (i) Keeping low values of $\Delta$ reduces the advantage that the opponent player can take in the next move; (ii) As observed in Fig.~\ref{fi:comp_2}~(c) and (d), the accuracy of the players is reduced if the values of $\Delta$ are low, hence the opponents are vulnerable to  make more mistakes. 

The reduction of $\avg{\Delta}$ and the two related effects are emerging properties of well-played games rather than the conscious choice of the players.
However, to introduce a model for the games, we can think of this as a strategy executed by good players. 
We can say that elite players, being better trained, can execute this strategy more effectively. 
In this way, we can explain that the curves of $\avg{\Delta}$ and $\sigma_{\Delta}$ of high-level players decrease as compared to lower-level players.
Random games in which players make moves without any criteria are equivalent to the lowest competitive level, hence as expected the curves increase.

\subsection{A Simple Model for Decision Making Process}

The aim of this section is to integrate the empirical information gathered in our work and propose a toy model to simulate the decision-making process in the game. 
It is important to clarify that the model does not attempt to capture all the complexity inherent in the process, but rather to show that there are simple mechanisms that can explain the emergence of the global statistics we observe.
The model is based on an agent that performs a walk in a complex tree. Starting from the root node and moving to the deepest level of the tree, the agent advances one node per time step.
At each node of the tree, the agent has to decide which branch to continue with. This mechanism emulates players deciding which move to make.
We propose that the number of branches in the tree follows a normal distribution, as is the case in the real tree (see Fig.~\ref{fi:arbol}~(b)). However, due to obvious computational limitations, we will use smaller trees than the real one, while maintaining approximately the relationship $\avg{\mu}/\sigma_{\mu}$ for greater structural similarity.
Each node of the tree is associated with a random value of $\Delta$. Upon reaching a node, the walkers use these values to decide which branch to continue based on the following algorithm:

\begin{enumerate}
    \item The walker reaches a node at depth $d$ with $\mu$ outgoing branches.
    
    \item It takes the values $\Delta_i$, $i=0,...,\mu$, from its neighbors at depth $d+1$,
    
       \begin{enumerate}
       \item If $d<d^{*}$, it assigns the nodes a probability $p_i\propto \Delta_i^{-1}$, and uses this probability to decide which node to continue its walk on.
       
       \item If $d>d^{*}$, it chooses the neighbor with the smallest value of $\Delta$ and continues its walk through that node.
       \end{enumerate}
    
    \item The procedure is repeated until reaching the deepest level of the tree, $d=D$.
\end{enumerate}

The main idea of the algorithm is that the walkers move with preferential attachment towards nodes with smaller values of $\Delta$, thus emulating the minimization process observed in the empirical data.
The value $d^*$ indicates a singular depth value in the tree. For $d<d^*$, we assume that the walker emulates a player who is not exactly sure which path to take, but their training allows them to identify some branches as better than others, so they move using the probabilities $p_i$.
On the other hand, for $d>d^*$, we assume that as the walker delves deeper into the tree, their analytical power increases, and they know which is the optimal path to take.
In this framework, we can relate $d^*$ to the competitive level. Given two walkers $i$ and $j$, if $d_i^*>d_j^*$, we say that walker $j$ is more skilled than $i$, because they can detect the optimal path at a smaller depth.
To incorporate possible mistakes by the players into the model, we introduce noise into the system. We define $\eta$ as the probability of a walker choosing a path completely at random.

\begin{figure}[t!]
\centering
\includegraphics[width=1.\textwidth]{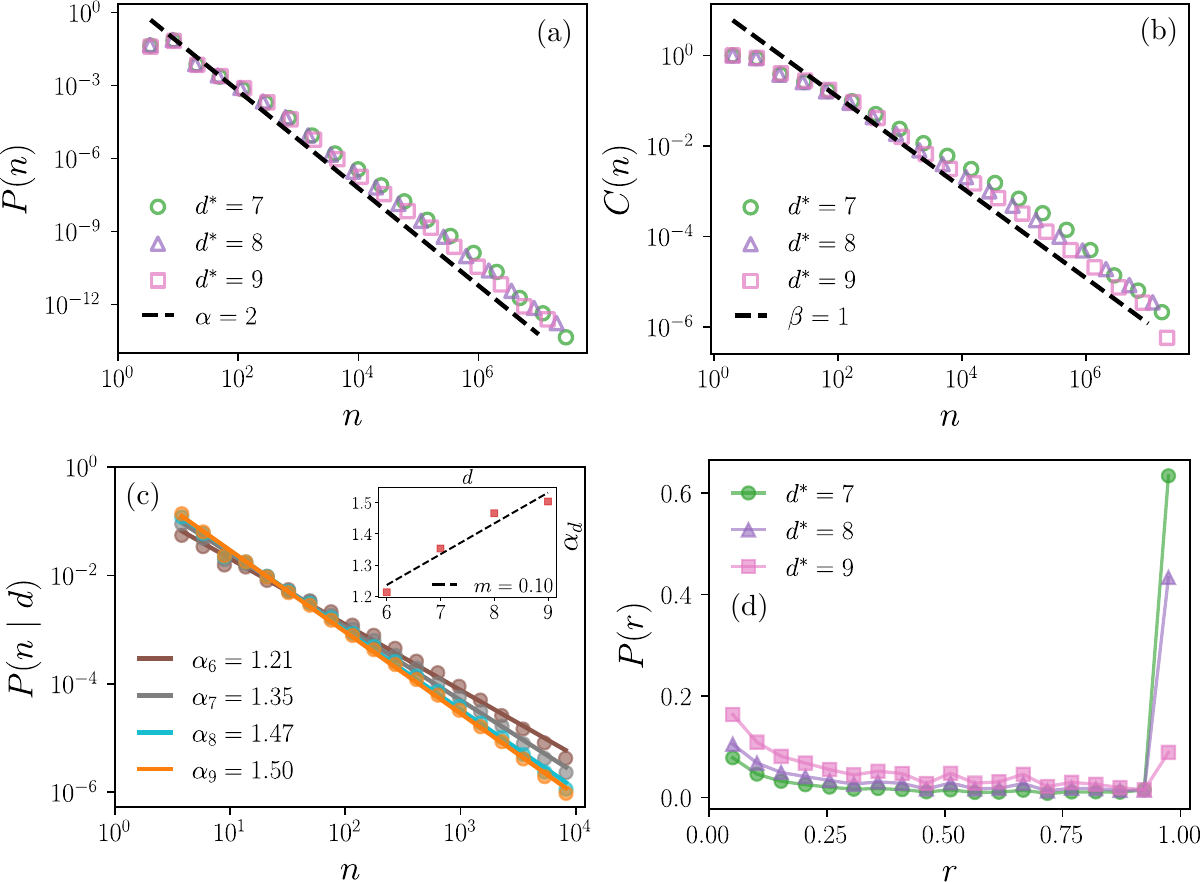}
\caption{
Model Results.
(a) Node popularity distribution \( P(n) \). Dashed lines show the theoretical distribution \( P(n) = n^{-2} \).
(b) Complementary cumulative distribution \( C(n) \). Dashed lines show the theoretical distribution \( C(n) = n^{-1} \).
(c) Distributions \( P(n|d) \) for the case \( d^* = 7 \), at depths \( d = 6, 7, 8 \), and \( 9 \). Solid lines show the curves associated with a nonlinear fit of the distributions with the expression \( P(n|d) = A~n^{-\alpha_d} \). The inset shows the relationship between the calculated \( \alpha_d \) values and depth.
(d) Ratio distribution \( P(r) \).
}
\label{fi:model}
\end{figure}

In this framework, we simulate walks on trees with nodes having a branch distribution $\mu \sim N(6,2)$. 
To set the values of $\Delta$ in the tree, we use the non-linear fit of the empirical probability distribution associated with the null model (see Fig.~\ref{fi:rw}~(a)), as it provides information about the landscape of values that a walker might encounter. Using the inverse transform technique in the stretched exponential case, we then generated the sampling to set the value of $\Delta$ at each node.
On the other hand, we set the noise level to a low value, $\eta=0.01$.
To study the effect of the $d^*$ values, we conduct three simulations with $d^*=7$, $8$, and $9$. In each case, we simulate $50$ million walks.
From the simulation results, we extracted the popularity of the nodes, and based on that, we calculated the distributions shown in Fig.~\ref{fi:model}.
In panels (a) and (b), we show the distributions $P(n)$ and $C(n)$. We observe that the curves follow the trend $P(n)\propto n^{-2}$ and $C(n)\propto n^{-1}$, which is consistent with what is observed in distributions associated with real games. 
In Fig.~\ref{fi:model}~(c), we show the distributions $P(n|d)$ for $d=6,7,8,$ and $9$. These curves were calculated for the results associated with $d^*=7$. We observe distributions with a heavy tail, with the slope increasing slightly as the depth increases. Using the maximum likelihood method, we fit the distributions and obtained the exponents $\alpha_d$ for each case. In the inset of the panel, we show that the relationship $\alpha_d$ vs. $d$ follows a similar increasing trend as observed in the real data.
In Fig.~\ref{fi:model}~(d), we show the distribution of ratios. We observe that around $r\approx1$, $P(r)$ shows a peak that decreases as the value of $d^*$ increases. 
Additionally, we see that at $d\approx0$, the probability value increases as the value of $r^*$ increases.

\begin{figure}[t!]
\centering
\includegraphics[width=1.\textwidth]{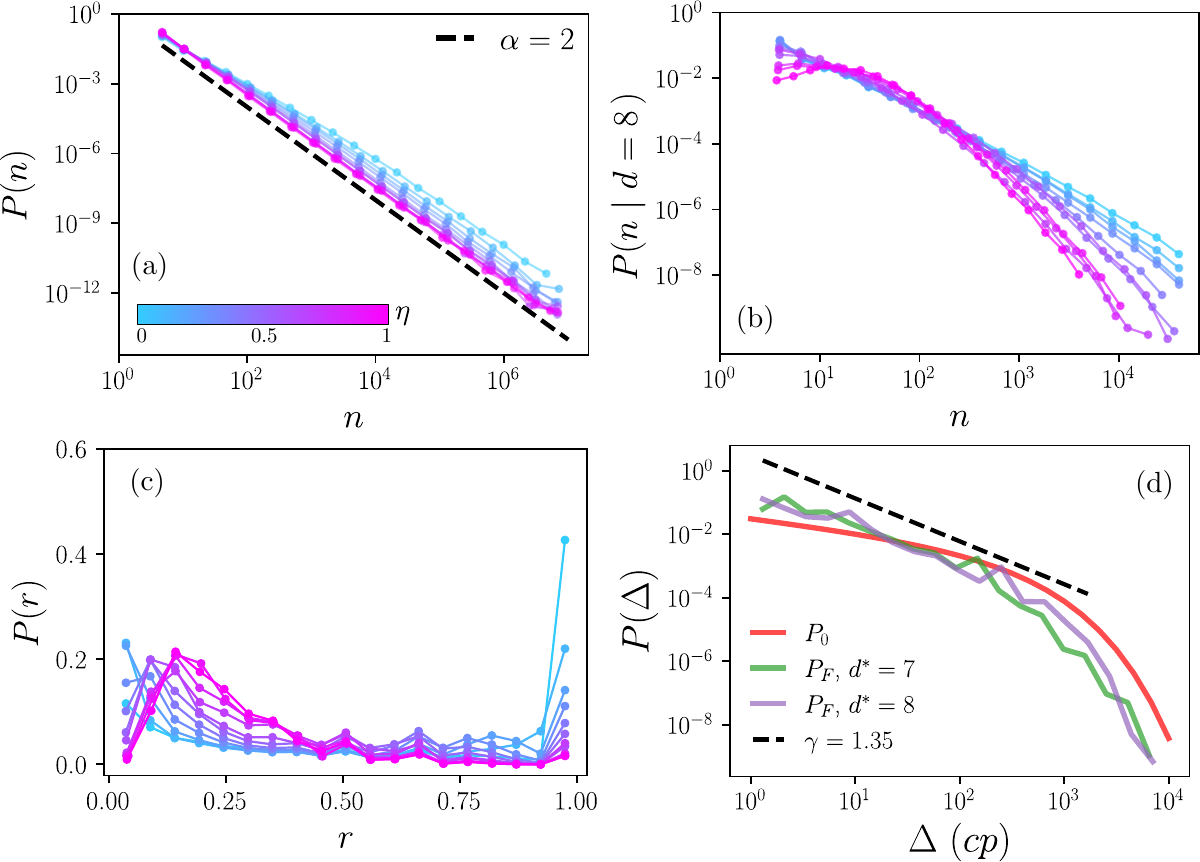}
\caption{
The curves in panels (a), (b) y (c) show the effect of the noise in the model for $d^*=8$.
(a) Node popularity distribution \( P(n) \),
(b) Node popularity distribution at depth 8, \( P(n|d=8) \),
(c) Ratio distribution \( P(r) \).
 The color bar relating the noise level to the curves is shown in panel (a).
(d) Comparison between the initial probability distribution of $\Delta$ values, $P_0$, set at the beginning of the simulation, and the probability distribution of $\Delta$ values explored by the agents during the walk, $P_F$.
}
\label{fi:ruido}
\end{figure}

To test the effect of noise, we conducted a study varying the value of $\eta$. 
For these simulations, we fixed the parameter $d^*=8$. The tree structure and the number of walkers were kept the same as in the previous simulations.
In Fig.~\ref{fi:ruido}~(a), (b), and (c), we show the distributions $P(n)$, $P(n|d=8)$, and $P(r)$, respectively.
The curves represent the results for $\eta=0.0, 0.1, ...1.0$, where the cyan color in the graph indicates low noise levels, while magenta indicates high levels.
Firstly, we observe that the noise level does not seem to significantly affect the exponent of the distribution $P(n)$ for large values of $n$. 
For high noise values, $\eta \to 1$, the curve closely resembles the theoretical one, $P(n) \propto n^{-2}$, which is consistent with a random walk on an approximately regular tree.
Secondly, we see that the distribution $P(n|d=8)$ becomes more homogeneous for high levels of noise, meaning that at the same level, we find similar popularity values.
Lastly, in the distribution $P(r)$, we see that noise destroys the peak at $r\approx1$ and creates another around $r\approx1/6$. This is related to what is observed in $P(n|d=8)$: noise makes it equally likely to go to any node, and since the average number of exits from a node is $\avg{\mu}=6$, when it is equally likely to take any path, the value of $r$ around $1/6$ increases.
Finally, in Fig.~\ref{fi:ruido}~(d), we compare the distribution of $\Delta$ values set as the initial condition, $P_0$, with the distribution of $\Delta$ values explored by the walkers during the simulations, $P_F$, for the cases $d^*=7,8$. As a reference, we show the theoretical curve $P(\Delta) \propto \Delta^{-\gamma}$ with $\gamma=1.35$, associated with the FIDE dataset, in dashed lines. We can see that the model approximately reproduces the process of minimizing $\Delta$ values observed empirically in Fig.~\ref{fi:rw}~(a).

As we can see, our toy model roughly captures the general aspects of the global emergent properties of the system. This by no means implies that our model is capable of capturing the total complexity of decision-making dynamics; rather, it suggests that despite being an extremely complex system, the mechanisms generating the observed global emergent properties can be approximated by relatively simple mechanisms based on structural parameters of the tree and the process of minimizing $\Delta$.

\section{Conclusions}

The main goal of our work was to shed light on the microscopic mechanisms that generate global emergent properties of the system. In this framework, we decided to focus our analysis on the players' decision-making process. To this end, we used the parameter $\Delta$ as a proxy, which indicates how crucial it is to make a good decision in a given position. We observed that as players progress through the game, they encounter positions with a wide range of $\Delta$ values, evidencing the complexity of the process. In this context, we noted statistically significant differences between groups of winners and losers, suggesting that a drop in the system's complexity could be associated with a drop in players' performance.
This is somewhat similar to what is observed in so-called living systems, such as in nervous system diseases that affect neuron interactions \cite{albano1996self}, or with the death of mitochondria within cells \cite{zamponi2018mitochondrial}. If complexity decreases in these systems, their functioning is severely impacted. The loss of the delicate balance between inhibition and promotion, cooperation and competition in a complex system leads to the appearance of something abnormal. In the game of chess, we observed this effect manifesting in the drop of players' competitive abilities.

On the other hand, by studying how well players choose the optimal move in a given position (accuracy), we found evidence that this improves in the presence of high $\Delta$ values. We also showed that this increase in accuracy is observed regardless of the players' competitive level or whether they belong to winning or losing groups. To complement this analysis, we conducted a comparative study between the empirical data and the results of a null model based on players making random legal moves. We observed that players tend to decide guided by a process of minimizing $\Delta$ values, and that this process improves with the players' competitive level. Note that this is consistent with our accuracy study. Given that players at any level become more accurate in the presence of positions with high $\Delta$ values, it is natural to think that, in the context of sporting competition, a winning strategy would be to make moves such that the opponent is left with positions having low $\Delta$ values. In this way the profit of the opponents is minimized and their probability of error is maximized.

Finally, using the set of empirical observations made in our work, we proposed a simple decision-making model. This model is based on agents walking in a complex tree where each node has a $\Delta$ value sampled from a power-law distribution. At each step, the agents have to decide which path to take, using a strategy that favors paths with nodes having small $\Delta$ values, attempting to emulate the minimization process observed in reality. We found that the simple mechanisms proposed in the model approximately generate the global emergent properties observed in the real system, demonstrating that the decision-making process, despite being extremely complex, can be explained using relatively simple evolution rules.

The framework, results, and analysis of the complexity of chess reported in this work can be useful for understanding  several
player's behaviors, and also to improve the interaction of players with engines.
Extensions of the statistic of critical points would have straightforward applications
among the following lines that are currently under study.
In the analysis of the behavior of players, Chowdhary et al.~\cite{chowdhary2023quantifying} found that skilled players can
be distinguished from the others based on their gaming behaviour where these differences appear from the very first moves of the game.
Experts specialize and repeat the same openings while beginners explore and diversify more.
Our results are in agreement with these findings and further specific analyses are
possible to provide a better understanding of this topic.
Another application of our work is in studying the popularity of opening lines. For instance, Lappo et al.~\cite{lappo2023cultural}
carried out an analysis of chess in the context of cultural evolution,  describing multiple cultural factors that affect move choice.
Also, our framework can be used to unveil risk patterns within chess games; complementing existing approaches to risk based on categorizing chess openings or entire games~\cite{carow2024time, dreber2013beauty}.
The relationship between chess and Artificial Intelligence (AI) is synergetic. An analysis of AlphaZero systems
suggests~\cite{mcgrath2022acquisition}  that it may be  possible to understand a neural network using human-defined chess concepts, even if it learns entirely on its own.
More specifically, using a machine learning model Tijhuis et al.~\cite{tijhuis2023predicting} showed that for predicting the level of a player the most important features to be considered during the match are both from theory, such as mobility at the end of the game and engine evaluation such as blunders and mistakes. Our analysis can provide new features to encode games and predict the level of expertise of players using AI.

\section{Data availability} 

The data used in this work can be requested from the first author (A. Chacoma: achacoma@df.uba.ar).

\section*{Acknowledgement} 

To the memory of our friends Oscar Cervantes and Carlos Vannicola, for teaching us about the beauty of the game.
This work was partially supported by CONICET under Grant No. PIP 112 20200 101100 and SeCyT-UNC (Argentina).

\newpage
\clearpage

\end{document}